\documentclass[conference]{IEEEtran}
\IEEEoverridecommandlockouts
\usepackage{cite}
\usepackage{amsmath,amssymb,amsfonts}
\usepackage{algorithmic}
\usepackage{graphicx}
\usepackage{textcomp}
\usepackage{xcolor}

\usepackage{blindtext}
\usepackage{xspace}
\usepackage{subcaption}
\usepackage{svg}
\usepackage{array}
\usepackage{graphicx}
\usepackage{mathtools}
\usepackage{hyperref}

\graphicspath{ {./figures/} }
\usepackage{anyfontsize}
\newcommand{\myfontsize}{\fontsize{5pt}{10pt}\selectfont}
\newcommand{\myfontsizesmall}{\fontsize{4pt}{10pt}\selectfont}

\def\BibTeX{{\rm B\kern-.05em{\sc i\kern-.025em b}\kern-.08em
    T\kern-.1667em\lower.7ex\hbox{E}\kern-.125emX}}

\makeatletter
\newcommand{\linebreakand}{%
  \end{@IEEEauthorhalign}
  \hfill\mbox{}\par
  \mbox{}\hfill\begin{@IEEEauthorhalign}
}
\makeatother

\begin{document}

\title{User Value in Modern Payment Platforms:\\A Graph Approach
\thanks{The research leading to these results has been funded by the European Union’s funded Project INFINITECH under Grant Agreement No. 856632 and by the SmartData@PoliTO center for Data Science technologies.}
}

\author{
\IEEEauthorblockN{Laura Arditti}
\IEEEauthorblockA{\textit{Larus Business Automation} \\
Venezia, Italy \\
laura.arditti@larus-ba.it}
\and
\IEEEauthorblockN{Martino Trevisan}
\IEEEauthorblockA{
\textit{University of Trieste}\\
Trieste, Italy \\
martino.trevisan@dia.units.it}
\and
\IEEEauthorblockN{Luca Vassio}
\IEEEauthorblockA{
\textit{Politecnico di Torino}\\
Torino, Italy \\
luca.vassio@polito.it}
\linebreakand 
\IEEEauthorblockN{
Alberto De Lazzari}
\IEEEauthorblockA{\textit{Larus Business Automation} \\
Venezia, Italy \\
alberto.delazzari@larus-ba.it}
\and
\IEEEauthorblockN{Alberto Danese}
\IEEEauthorblockA{\textit{Nexi} \\
Milano, Italy \\
alberto.danese@nexigroup.com}
}

\maketitle

\begin{abstract}

Payment platforms have significantly evolved in recent years to keep pace with the proliferation of online and cashless payments. These platforms are increasingly aligned with online social networks, allowing users to interact with each other and transfer small amounts of money in a Peer-to-Peer fashion. This poses new challenges for analysing payment data, as traditional methods are only user-centric or business-centric and neglect the network users build during the interaction. This paper proposes a first methodology for measuring user value in modern payment platforms. We combine quantitative user-centric metrics with an analysis of the graph created by users' activities and its topological features inspired by the evolution of opinions in social networks. We showcase our approach using a dataset from a large operational payment platform and show how it can support business decisions and marketing campaign design, e.g., by targeting specific users.

\end{abstract}


\section{Introduction}
\label{sec:intro}

% \begin{enumerate}
%     \item Context: online payments are widespread
%     \item Platforms are more and more community-oriented 
%     \item To expand the user base and increase engagement, it is necessary to incentivize users 
%     \item Our goal is a methodology to support decision making when it comes to distribute incentives 
%     \item User-centric vs network perspective
%     \item Real world data from YAP
% \end{enumerate}
% \hrulefill

Various forms of online payments have emerged in the landscape of payment systems to support the rapid growth of e-commerce. They are already widespread, often surpassing more traditional payment methods~\cite{sumanjeet2009emergence, mukherjee2017commece}. Online payments provide users with a fast, direct, convenient, and secure way to access their funds and conduct transactions, which is among the main reasons for the success of e-payment technologies in terms of their widespread adoption~\cite{khan2017compendious}. In particular, mobile payment services are growing in popularity as they keep pace with the shift toward massive mobile internet access~\cite{bezovski2016future, su2018users, church2007mobile}. At the same time, mobile payment platforms are becoming more and more community-oriented: peer-to-peer (P2P) payment platforms~\cite{caceres2020peer} offer users new ways to interact with each other to the point where they are evolving into a new form of online social network~\cite{acker2018venmo}.
 
Great efforts are made to ensure the success of a mobile payment solution. In particular, companies aim to expand the user base and increase user engagement. Common approaches to these tasks focus on designing incentives and reward mechanisms to keep existing users active and attract new valuable users~\cite{staykova2015race, ho2022promoting}. 
When developing strategies to improve the quality of the user base, one of the biggest challenges is understanding users' value and measuring it operationally, as this is critical for business decisions such as designing marketing campaigns. Classical user-centric approaches measure the value of users only based on their individual activities. However, they can hardly capture their quality in an environment characterized by a network of interactions, which poses severe limitations to data analytics. Indeed, there is a lack of methodologies to analyze user behavior in this new generation of payment platforms, where interactions between users play an essential role. It is therefore crucial to develop effective metrics to guide business decisions and support the work of marketing departments.  Mobile payment platforms can be seen as constantly expanding networks through a ``member-get-member'' mechanism. Hence a good measure of users' value must capture not only their spending habits, but also how each user contributes to the creation of a high-quality network.

In this paper, we propose a methodology that models payment platforms as networks and exploits their structure to guide business strategies. We overcome the limitation of current metrics for quantifying user value in payment systems by leveraging relationships between users. Inspired by graph mining methods, our approach combines user-centric features with topological features extracted from the payment graph. We also present a practical way to compute this new metric using an iterative graph algorithm. To the best of our knowledge, we are the first to propose a practical method for measuring user value in current payment platforms, which are characterized not only by traditional activities (purchases from merchants) but also by P2P interactions between users.

We present our methodology using a dataset of a real-world payment network. More specifically, we present the business case of YAP, Nexi's mobile payment platform, which allows users to perform payments at physical and online merchants and to exchange money with each other. We applied our methodology to analyze the YAP platform's community, with the goal of supporting business decisions. %over the last three years. 
Our results show that the approach is a practical tool to support marketing
campaigns and, more in general, business decisions.

%After an in-app enrollment, YAP is active as a MasterCard prepaid and accepted by merchants both worldwide and online. Yet, YAP is more than just a prepaid card, it is a service that allows its users to interact with each other in multiple ways: splitting bills, sending or asking money, inviting new friends. These mechanisms characterize YAP as a P2P payment platform and endow YAP data with an underlying graph structure, which can be leveraged with our methodology.

\section{Dataset and its graph representation}
\label{sec:dataset}

% \begin{enumerate}
%     \item Our dataset comes from a real payment network.
%     \item It includes many users and merchants.
%     \item Users and merchants are characterized by some features.
%     \item Thee three type of relations between nodes.
%     \begin{itemize}
%         \item Users buy on merchants and form a graph.
%         \item Users can send payments to other users.
%         \item Users can invite new users in the platform.
%     \end{itemize}
%     \item Characterization and examples
% \end{enumerate}
% \hrulefill

To develop and evaluate our methodology, we take a data-driven approach and use as a reference a dataset collected from an operational payment platform. 
The dataset comes from the Italian app YAP\footnote{\url{https://www.yap-app.it}}, a payment platform provided by Nexi\footnote{\url{https://www.nexigroup.com/en/}}, one of the biggest European players in digital payments. YAP is based on a mobile application linked to a prepaid card (accepted by online and physical stores) that also allows its customers to exchange money with friends and contacts without fees.
In this paper, we use data from the production databases of YAP, which include a set of transactions for the years 2019, 2020 and 2021, as well as metadata about users and merchants.

\begin{figure}[t]
    \myfontsizesmall{\includesvg[width=8cm]{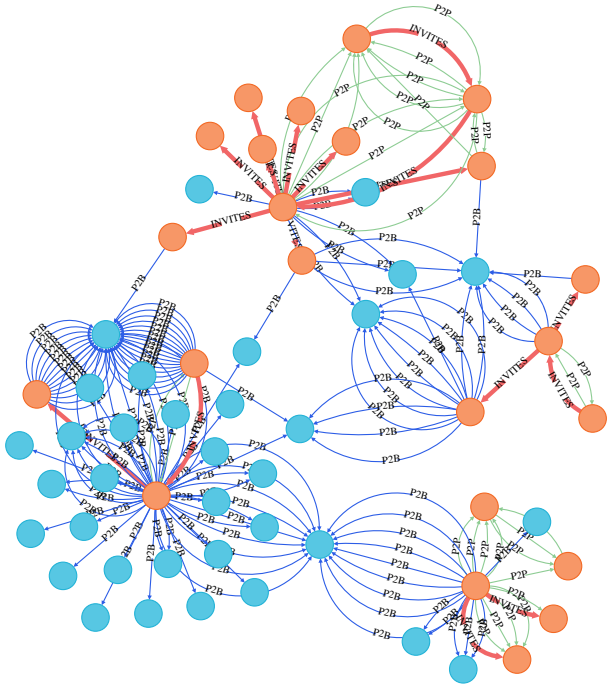}}
    \centering
    \caption{A small portion of the dataset. Users and merchants are represented as orange and blue nodes, respectively. Invitations among users are represented with bold orange links, P2B transactions correspond to blue links, P2P money transfers are displayed in green.}
    \label{fig:dataset} 
\end{figure}
    
The dataset can be naturally represented in terms of a heterogeneous graph, since there are entities that are related to each other. In particular, we have identified three types of relationships that reflect the three main types of interactions between users and merchants.
\begin{enumerate}
    \item Users are connected to merchants by ``P2B'' relationships, representing monetary transactions characterized by their date, amount and channel, which may be online (i.e., e-shops) or offline (i.e., physical stores).
    \item Users may transfer money to other users. This kind of interaction is represented by ``P2P'' relationships among users, which are characterized by their date and amount.
    \item Finally, users may invite new users to join the platform. This results in ``Invite'' relationships, whose tail and head nodes correspond to users sending and accepting the invitation, respectively. Note that we only model invitations that resulted in the acquisition of a new users.
\end{enumerate}
These relationships are characterized by a timestamp. Hence we have a dynamic graph, with edges appearing and disappearing over different time windows.

We sketch a small portion on this heterogeneous graph in Figure~\ref{fig:dataset}, where users and merchants are connected with three types of edges. For privacy reasons, we anonymize the dataset by removing personally identifiable information. As a result, users and merchants are identified by unique numeric identifiers. Each user is associated with some personal details (age, gender, place of residence, occupation), while merchants are characterized by a category indicating the type of activity and the province of their retail store.

We store our dataset in the graph database Neo4j\footnote{\url{https://neo4j.com}}, which provides a native representation of graph data, so we could efficiently traverse the graph, query it for patterns and visualize the resulting information. The dataset is quite large and includes a number of nodes in the range $(10^6,10^7)$ and a number of relationships in the range $(10^7,10^8)$.\footnote{We cannot disclose the exact numbers and ranges as they represent trade secrets.} % Range concordati con Nexi, da non modificare

\begin{figure}[t]
    \myfontsizesmall{\includesvg[width=8cm]{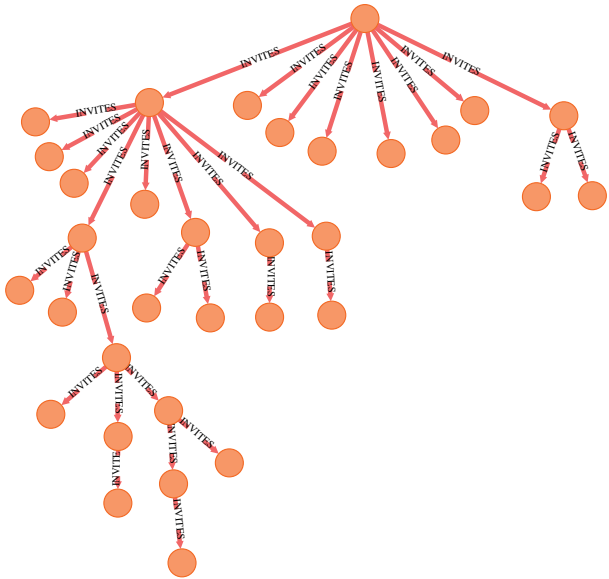}}
    \centering
    \caption{Example of WCC showing the invitation spreading process triggered by the invitations sent by the root top user.}
    \label{fig:invitation-spreading}
\end{figure}

\begin{figure}[t]
    \includegraphics[width=\columnwidth]{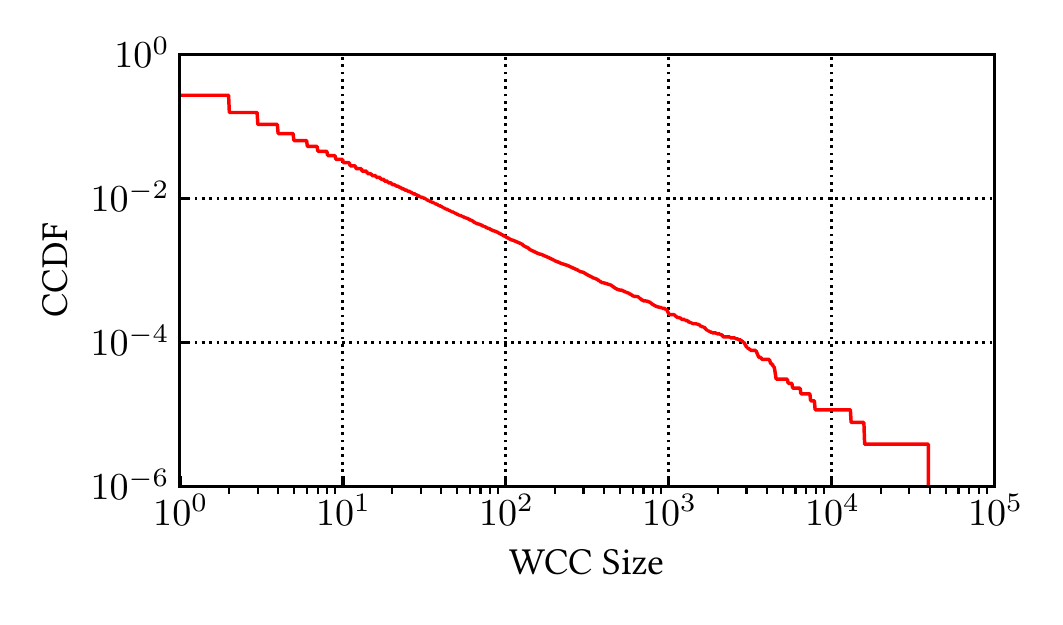}
    \centering
    \caption{Distribution of weakly connected component sizes for the invitation network $\mathcal G$. Vertical axis is in log-scale.}
    \label{fig:wcc-size}
\end{figure}

\begin{figure}[t]
    \includegraphics[width=\columnwidth]{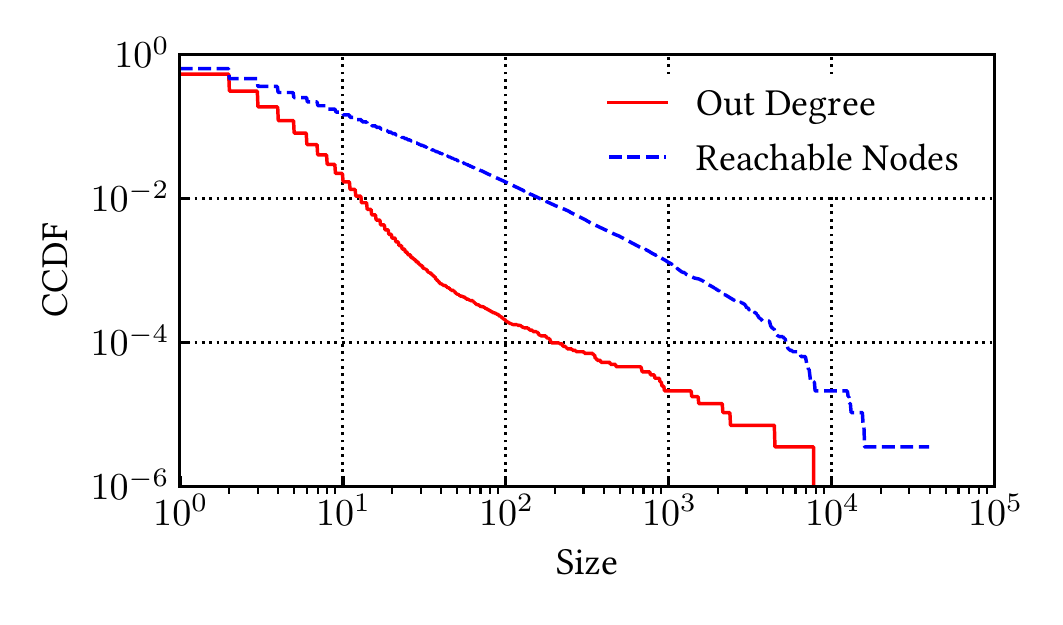}
    \centering
    \caption{Out degree and reachable nodes CCDF for nodes in the invitation network $\mathcal G$. Vertical axis is in log-scale.}
    \label{fig:out-degree}
\end{figure}

For our methodology, the ``Invitation Network'' plays a relevant role. It simply represents the network of users connected by the ``Invite''  relationships. Formally, we define it as the subgraph $\mathcal G = (\mathcal V, \mathcal E)$ of our dataset comprising all users $\mathcal V$ and the invitation relationships among them.\footnote{Merchants cannot invite neither users or other merchants to the platform.} The edges $\mathcal E$ then represent the ``Invite'' relationships among couples of nodes $(u,v) \in \mathcal E \subset \mathcal V \times \mathcal V$. The Invitation Network $\mathcal G$ plays a key role in the development of our methodology, as it captures the temporal evolution of the YAP network in terms of new users acquired through accepted invitations. We therefore briefly characterize its main topological features.
%Denote with $\mathcal N_u$ the out-neighborhood of node $u$ in $\mathcal G$, defined by
% \begin{equation}
%     \mathcal N_u = \left\lbrace v \in \mathcal V | (u,v) \in \mathcal E \right\rbrace, 
% \end{equation}
% i.e., the set of users that received and accepted an invitation from user $u$.
% Based on this, the out-degree $d_u$ of node $u$ in $\mathcal G$ is defined as the number of its out-neighbors, i.e. the number of other users that entered the YAP network by accepting an invitation from $u$:
% \begin{equation}
%     d_u = |\mathcal N_u| \,.
% \end{equation}
First, we note that the invitation graph has a special structure: $\mathcal G$ is a forest, i.e., each weakly connected component (WCC) of $\mathcal G$ is a directed tree, since each user can send many invitations but he can accept only one. An example of a WCC from the dataset is shown in Figure~\ref{fig:invitation-spreading}. The top user sent several invitations, 8 of which were accepted. Some users in turn invited other users, forming a WCC with a total of 34 users. The size of the WCCs varies from small single-user or two-user components (none or a single accepted invitation) to subgraphs with hundreds of users. In Figure~\ref{fig:wcc-size}, we show the distribution of WCC size in terms of a complementary cumulative distribution function (CCDF). The logarithmic axes highlight the presence of a huge component that includes nearly $4 \times 10^5$ users. The remaining WCCs are smaller and have different shapes, with some resembling an invitation chain (each user sent a single invitation), while others contain spreader users (i.e., users who sent a large number of invitations). We quantify this in Figure~\ref{fig:out-degree}, where we show the distribution of out degree of nodes. Most users sent a handful of invitations, while a limited group (a  dozed of individuals) sent several thousand -- note the tail of the distribution in the bottom right of the figure. The median out degree is 0, as only 28\% of users sent at least one invitation. In Figure~\ref{fig:out-degree} we also show the distribution of reachable users starting from a node, i.e., the size of the largest sub-tree of $\mathcal G$ for which a node is the root,  representing the cascade of invited customers. The significant difference between the two distributions highlights that it is relevant to consider the impact of users on the whole network, not just on their one-hop neighborhood. 
 
\section{Methodology for assessing user value}
\label{sec:metho}

% \begin{enumerate}
%     \item The goal and blocks of our methodology.
%     \item The intrinsic score, inspired by the RFM model.
%     \item The network score, inspired by the theoretical works.
% \end{enumerate}
% \hrulefill

The goal of our methodology is to provide a practical and effective means for measuring the value of a user in a payment network. To this end, we compute the value of each user by means of a two-step process. First, we obtain the \emph{Intrinsic Value} of a user, which captures her value based on quantitative characteristics (e.g., the number of purchases). Then, we combine the Intrinsic Value and the graph structure of the Invitation Network to compute the final \emph{Value}. This measures the contribution of a user to the platform, taking into account the network of users characterized by relationships among them.

\subsection{Intrinsic Value}
\label{sec:intrinsic}

We start with the definition of \emph{Intrinsic Value}, whose goal is to represent the value of a user seen as an isolated entity. It is a score that measures the level of activity of a given user and combines her individual characteristics using an extended continuous RFM model~\cite{fader2005rfm}. Similar to classical RFM models, our score captures the spending habits of users (in terms of transactions recency, frequency and monetary value, hence the name RFM). We extend the RFM model by also measuring the users' activity level in the context of the invitation mechanism and we adapt it to a continuous setting, by deriving fine-grained information about users' characteristics instead of performing a classification task. In detail, the intrinsic value $I(u)$ of a users $u$ is obtained as the following product of four distinct factors
\begin{equation}\label{eq:intrinsic}
    I(u) =  M(u) \sigma_{R}(R(u)) \sigma_{F}(F(u)) \sigma_{E}(E(u)) .
\end{equation}
The first factor, $M(u)$, is a measure of the user's spending, given by a linear combination of their total online and offline transactions amounts whose weights represent the relative value of the two channels. $R(u)$ measures transaction recency, i.e., the duration (in months) since the last transaction performed by the user, while $F(u)$ takes into account the weekly transaction frequency.
Finally, the expansion term $E(u)$ measures the ability of the user to expand the YAP network, based on the number of invitations sent and accepted (i.e., their degree in the invitation network $\mathcal G$).
All factors except for $M(u)$ 
range in the interval  $(a,b)=(1/2,2)$ and
undergo generalized sigmoid functions designed to reward or penalize good or bad values for the corresponding quantity, according to the following formula:
\begin{equation}
    \sigma(x) = a + \frac{b-a}{1 + \exp \left( - s(x-c) \right)} \,,
\end{equation}
where for each factor appearing in \eqref{eq:intrinsic} the parameters  $c$ and $s$ are chosen to obtain desirable center and saturation points (where $\sigma(x) = \frac{a+b}{2}$ and $\sigma(x) \approx b$), selected according to business knowledge or based on the corresponding feature's distribution. In our experiment, we fix their value to $c=2$, $s \approx -1.3$ for the recency factor, $c=0.25$, $s \approx 3.5$ for the frequency factor and $c=1.5$, $s \approx 1$ for the expansion factor, in order to obtain saturation at $x=0$, $x=1$ and $x=4$ respectively. 
%Hence the factors range in the interval  $(a,b)=(1/2,2)$.
The resulting intrinsic value can be treated as a monetary value, and in our case is measured in euros. 
Notice that the intrinsic value $I$ is by definition a non-negative quantity. If a user $u$ is not active on the P2B side, i.e., if he has never performed any monetary transaction, his intrinsic value is vanishing, $I(u)=0$.

\begin{figure}[t]
    \centering
    \begin{subfigure}[b]{0.5\textwidth}
        \centering
        \myfontsize{\includesvg[width=0.45\textwidth]{network-value-low-nodec.svg}}
        \caption{A user with high intrinsic value and little network value.}
        \label{fig:network-low}
    \end{subfigure}
    \begin{subfigure}[b]{0.48\textwidth}
        \centering
        \myfontsize{\includesvg[width=\textwidth]{network-value-high-nodec.svg}}
        \caption{A user with high network value but lower intrinsic value compared to that of top user in Figure \ref{fig:network-low}.}
        \label{fig:network-high}
    \end{subfigure}
    \caption{Comparison between two users with different intrinsic and different network value. Intrinsic value of users is reported within nodes. }
    \label{fig:network-high-vs-low}
\end{figure}

Although it summarizes the salient characteristics of a user, the Intrinsic Value alone is not a complete measure of the value she brings to the payment platform. Indeed, it does not allow us to distinguish between two users with similar spending habits but a very different impact on the platform's development (and expansion). We show an example of this situation in Figure~\ref{fig:network-high-vs-low}. In Figure~\ref{fig:network-low}, the root user (the top one) has a high Intrinsic Value (shown as the value inside the node), i.e., she has good spending habits. However, she invited only three users, who are of poor quality and did not propagate the invitation process any further. Conversely, in Figure~\ref{fig:network-high} the root user has a lower Intrinsic Value but contributes significantly to the expansion of the payment platform. She invited three high quality new users, who started a cascading invitation process that spread widely. These two situations, shown in the two figures, are very different but indistinguishable under the lens of Intrinsic Value. Indeed, the value of a given user cannot be uniquely estimated using quantitative activity metrics (i.e., the intrinsic value), but it is codified in the way users relate to each other in a network of invitations. Thus, to successfully address the problem, we need to take this structure into account.

\subsection{Network Value}
\label{sec:network}

Following the above considerations, we start from the Intrinsic Value to compute a more effective score that we simply call \emph{Value} of a user. Its goal is to capture the contribution of a given user in the growth of the payment network. We here discuss the main technical details. Starting from the Invitation Network $\mathcal G = (\mathcal V, \mathcal E)$, the value $V$ of users is defined as the solution to the following system of equations:
\begin{equation}\label{eq:value-measure}
    V(u) = \alpha \underbrace{\left( \frac{1}{|\mathcal N_u|} \sum_{v \in \mathcal N_u} V(v) \right)}_
    {\mathclap{N(u)}} + I(u) , \quad \forall u \in \mathcal V
\end{equation}
where $\mathcal N_u$ are the neighbor nodes of $u$ while the scalar $\alpha \in (0,1)$ measures the relative importance of connections with respect to intrinsic value in determining users' overall value and acts as a damping factor, so that the effect of a user $v$ on the value of $u$ decays exponentially with their distance in $\mathcal G$. $\alpha$ is a parameter of our model that must be tuned to reflect the importance of user-base expansion with respect to monetary profit from a business perspective. In our experiments, $\alpha$ was set to $0.85$. For clarity, we call \emph{Network Value} $N(u)$ the first term of the sum in \eqref{eq:value-measure}. %, that separately quantifies the network component of Value $V(u)$ 
Notice that, by definition, the Value $V$ is a non-negative quantity, which is vanishing for a user $u$ that has never performed any monetary transaction ($I(u)=0)$ if he has never sent any successful invitations or if all users he invited have vanishing value (so that also $N(u)=0$). 

Value $V(u)$ combines Intrinsic Value with the structure of the invitation network in such a way that each user is rewarded for inviting other users who either have good spending habits or continue to grow the network. The idea at the base of our definition reflects the expansion mechanism of the payment network. When a user invites new members, they may propagate the invitation by doing the same. The first invitation starts a cascading process that eventually leads to the expansion of the network. Our graph-based measure combines the intrinsic value of each user with the quality of the community they have created by starting the expansion process.

The value of users can be computed iteratively via a custom graph algorithm that elaborates users' data leveraging the structure of connections. More precisely, we first compute the intrinsic value $I$ of users, we set $V_0 \equiv 0$ (so that the initial estimate of users' value coincides with their intrinsic value $V_1 = I$) and then, for each iteration $t = 1, 2, \ldots$, we compute the step-$t$ estimate of users' value $V_t$ as 
\begin{equation}\label{eq:algo-iteration}
    V_{t}(u) = \alpha \left( \frac{1}{|\mathcal N_u|} \sum_{v \in \mathcal N_u} V_{t-1}(v) \right) + I(u) .
\end{equation}
In designing our algorithm we were inspired by the literature on averaging processes that are used to model the evolution of opinions and conventions in social networks~\cite{friedkin1990social}. 
We create a correspondence between the two settings (detailed in Appendix \ref{sec:app}) that guarantees convergence of our algorithm \eqref{eq:algo-iteration} to the solution of \eqref{eq:value-measure} for generic graph topologies. 

Note that, due to the specific forest structure of the network of invitations, this schema results in the propagation of users' value from leaf users upwards. As a result, we can bound the number of iterations needed for convergence with the size of the largest connected component of $\mathcal G$, which is finite.

We implement the computation of users' Value within the message passing framework of the Neo4j Pregel API.\footnote{\url{https://neo4j.com/docs/graph-data-science/current/algorithms/pregel-api/}} In this framework nodes possess a piece of information, stored as internal state. They can exchange information by sending messages through links, and they can process the received information to update their internal state. In this way, a distributed scheme can be defined in which each node iteratively computes its own Value, starting from an initial state that matches its own Intrinsic Value. In our experiments, running this algorithm on a graph with millions of nodes takes less than one minute using a commodity Linux server.

\section{Results}
\label{sec:results}

% \begin{enumerate}
%     \item Intrinsic score: distribution
%     \item Network score: examples and distribution. Contrasting users with low and high scores, by age, gender, region, etc.
%     \item Intrinsic vs network score
%     \item The temporal dimension
%     \item Use Case A: evaluating the performance of campaigns (not only in terms of new users, but also in terms of their value).
%     \item Use Case B: comparing P2P user habits with their value.
% \end{enumerate}
% \hrulefill

In this section, we illustrate the Value we obtain for users in the YAP platform, using our dataset described in Section~\ref{sec:dataset}. We then discuss the applications of user Value for marketing and business.

\subsection{Intrinsic and Network Values}
\label{sec:res-score}

\begin{figure}[t]
    \includegraphics[width=\columnwidth]{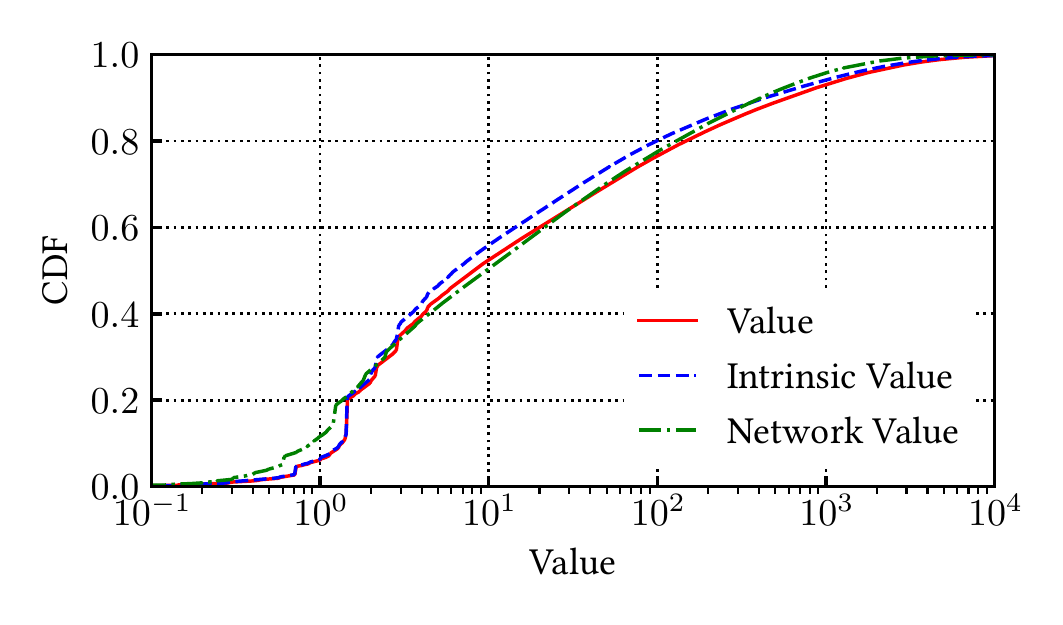}
    \centering
    \caption{Cumulative distribution function (CDF) of Value (overall, Intrinsic and Network). Notice logarithmic $x$-axis.}
    \label{fig:value-cdf}
\end{figure}

\begin{figure}[t]
    \includegraphics[width=\columnwidth]{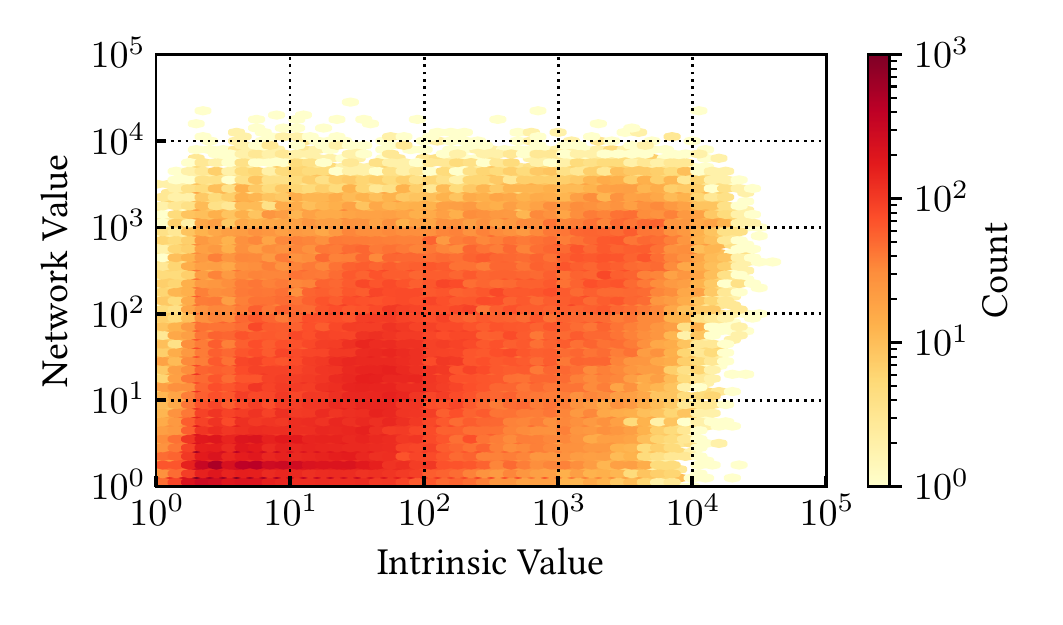}
    \centering
    \caption{Comparison of Intrinsic and Network Value of users. They are only 0.22 Pearson correlated.}
    \label{fig:intrinsic-vs-network}
\end{figure}

% \begin{table}[t]
%     \centering
%     \small
%     \begin{tabular}{ |m{1.8cm}|>{\centering\arraybackslash}m{1.7cm}>{\centering\arraybackslash}m{1.7cm}>{\centering\arraybackslash}m{1.7cm}| } 
%         \hline
%         & Intrinsic value & Network value & Value \\ 
%         \hline
%         Intrinsic value & $-$ & 0.22 & 0.95 \\ 
%         Network value & 0.22 & $-$ & 0.50 \\ 
%         Value & 0.95 & 0.50 & $-$ \\ 
%         \hline
%     \end{tabular}
%     \caption{Pearson correlation coefficients among Value components.}
%     \label{tab:pearson-corr}
% \end{table}

We first show the distribution of user Value, offering a break down for the Intrinsic and Network Value  in Figure~\ref{fig:value-cdf}. Notice that, following \eqref{eq:value-measure}, $N(u)$ directly measures the \emph{Network Value} of a user, that is, we are able to quantify the role of connections in determining each user’s value, also referred to as ``network effect''. %More precisely, the network value of users, $N(u)$, is computed from \eqref{eq:value-measure}.
% is formally defined as
% \begin{equation}\label{eq:network-value}
%     N(u) = V(u) - I(u).
% \end{equation}
Figure~\ref{fig:value-cdf} shows that the three curves span almost five orders of magnitude -- notice the logarithmic $x$-axis. Overall the curve have a  similar shape, with median around $10$. Less than $10\%$ of users have any Value component above $1\,000$. This is intentional as, by design, we want to have a measure that allows one to pinpoint users providing large benefits to the payment platform in terms of activity (Intrinsic Value) and ability to involve other users (Network Value). For instance, notice that only $0.18\%$ ($0.03\%$) of users have Intrinsic (Network) Value above $10\,000$. Overall,  the Value is far from being equally split among users. The top-10\% users hold 93\% of the total Value, while the Gini Index of the distribution is $0.89$, hinting great inequality.\footnote{The Gini Index or Gini Coefficient is a measure of dispersion (or inequality) in a distribution of samples.}

Intrinsic and Network Values have different goals. The former quantifies users' activity level, inspired by the RFM model~\cite{fader2005rfm}, widely adopted in  marketing. Conversely, Network Value of users is defined by the value of other users that are directly and indirectly connected to them by means of invitations. As exemplified in Figure~\ref{fig:invitation-spreading}, users’ invitations spread in a cascading process that expands the payment network, and we can identify a whole community of users that entered the network as the direct and indirect result of the actions of a single root user. The quality of such user’s community defines their network value. Notice that in defining the network value, quality is rewarded over quantity: it is not about the number of invited users, but the portion of them which is valuable. We compare the two Value components in Figure~\ref{fig:intrinsic-vs-network}, where we report the HexBin plot of the two quantities.\footnote{A HexBin plot is a variation of heatmap, where cells have a hexagonal shape.} We first observe a portion of users having vanishing Intrinsic and Network Value, corresponding to inactive users. Then, we notice many users with high Network Value but low Intrinsic Value (top left of the figure): these are valuable users that we miss if we only look at their individual activity. The opposite case -- users with high Intrinsic and low Network Value --, conversely, is slightly less frequent (bottom right of the figure). Overall, the main diagonal in the picture appears to be dense, as the two values are not completely independent. Indeed, their Pearson correlation coefficient is $0.22$, hinting positive but weak correlation. %As summarized in Table~\ref{tab:pearson-corr},
The overall Value is $0.95$ ($0.50$) correlated with the Intrinsic (Network) Value. This is expected, as a result of the high value set for the damping factor $\alpha$.

\subsection{Temporal Evolution}
\label{sec:res-temporal}

\begin{figure}
    \includegraphics[width=\columnwidth]{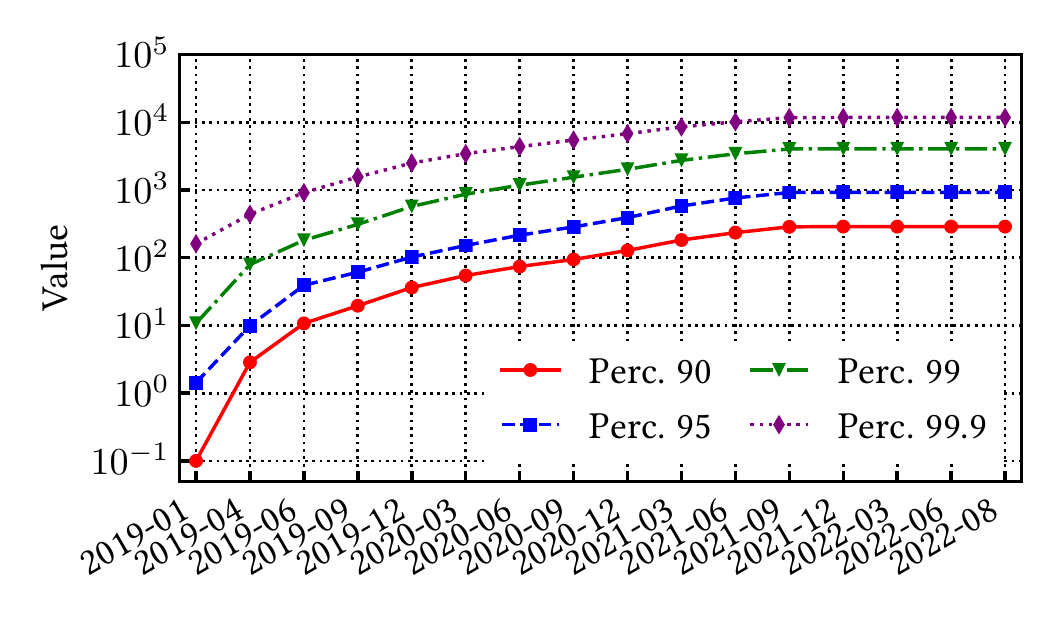}
    \centering
    \caption{Various percentiles of user value over time, from January 2019 to December 2021. }
    \label{fig:value-history}
\end{figure}

We now study the evolution of users' Value over time to understand how it varies when new users join the platform and run some activity (e.g., make payments or invite other users). To this end, we compute both Intrinsic and Network Value at different time instants, thus taking into account only transactions and invitations that already took place up to that time.
 At enrollment time, all users have vanishing Value. Users' value starts increasing when users start their transaction activity or when users they invited start to produce value. Note that neither the Intrinsic nor the Network Value are necessarily monotone increasing functions of time (and thus, as a consequence, neither is their sum, the Value). The Intrinsic Value $I(u)$ of a user may decrease over time if the user reduces or stops his transaction activity, so that the frequency and recency factors decrease. On the other hand, the 
 Network Value $N(u)$ of a user may be reduced if the value of other users he invited decrease or if he invites additional users with low value, which decrease the average quality of his out-neighborhood.
We perform a temporal analysis considering a series of expanding windows and we observe the evolution of users' Value over three years, starting from January $2019$ (shortly after the product was launched) up to December $2021$. We compute the users' Value every three months and show various percentiles of its distribution in Figure \ref{fig:value-history}. We note that the distribution spreads as time moves forward to represent the progressive acquisition of new users and the growth of the total value brought by users to the platform. All curves in the Figure exhibit an increasing trend, meaning that Value tends to increase over time. This increase is expected as, since its launch, the payment platform acquired a large number of new users, who, in turn, invited other people to join. The median Value passed from 0 in early 2019 to 2.14 in December 2021. Interesting is the tail of the distribution. There, we find users with very high value -- i.e.,  those to which marketing campaigns should give special attention. The top 1\% of users increased their Value from hundreds to several tens of thousands. Such a temporal analysis has proved helpful in the payment platform's business strategies. Indeed, it has enabled data-driven decision-making over the period, allowing identifying events that lead to a significant increase or decrease in the value of users. This gives a solid intuition on how to control the evolution of the user base.

\section{Discussion}

\subsection{Applications to Marketing and Business}

\begin{figure}[t]
    \includegraphics[width=\columnwidth]{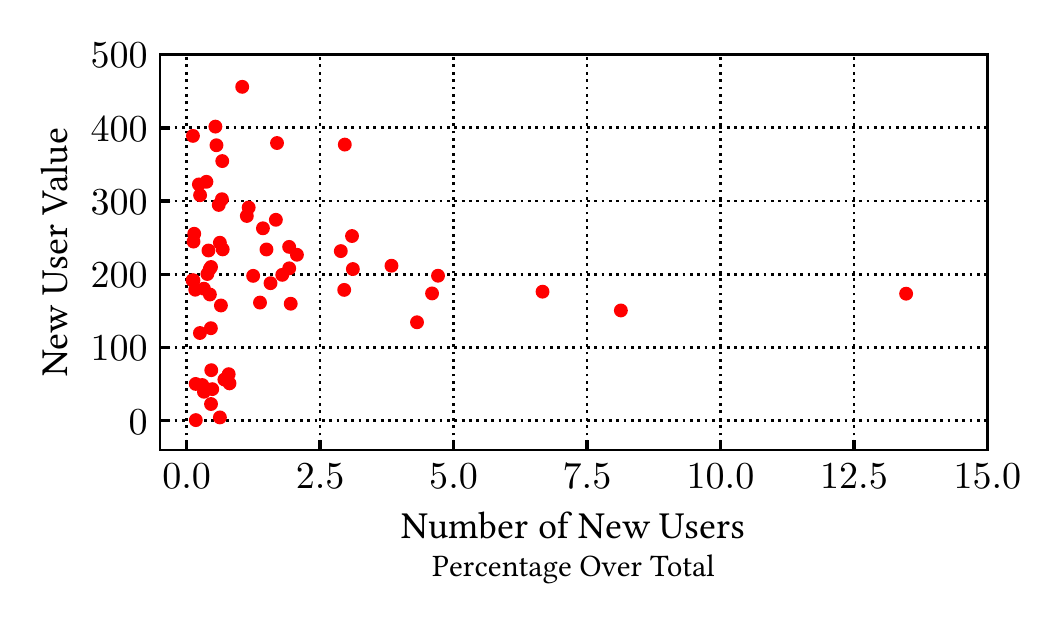}
    \centering
    \caption{Scatter plot showing the number of new users acquired on different marketing campaigns compared with their average Value.}
    \label{fig:campaigns}
\end{figure}

The adoption of this methodology to compute users' Value produced descriptive and prescriptive results, thus impacting both the understanding and operations of the payment platform business team. On the descriptive side, the Value allows answering business questions, such as: what marketing campaigns generate the most reward for the platform? Indeed, this graph based measure allows to effectively measure the profit generated by marketing campaigns. Instead of just counting the number of acquired users, we can measure the Value of new users enrolled during a campaign. This allows identifying what campaigns are able to attract more valuable users, which in turn allows understanding why such campaigns are more effective than others. As an example, in Figure~\ref{fig:campaigns} we compare the number of users acquired per marketing campaign with the average value of users these users for 63 marketing campaigns launched between 2019 and 2021. Thus, we can investigate what campaigns yield the best return of investment by comparing the value of newly acquired users with the amount invested in their acquisition. For instance, the points at the bottom left of the picture represent campaigns that attracted a limited number of new users, but with a high Value.\footnote{The Value of acquired users is computed at the end of the dataset, thus including the activity of these new users.} Conversely, the three right-most points refer to campaigns leading to the acquisition of a (relatively) large set of new users. However, their Value (as computed several months after the campaigns) is not as high as the former example.

\begin{figure}[t]
    \includegraphics[width=\columnwidth]{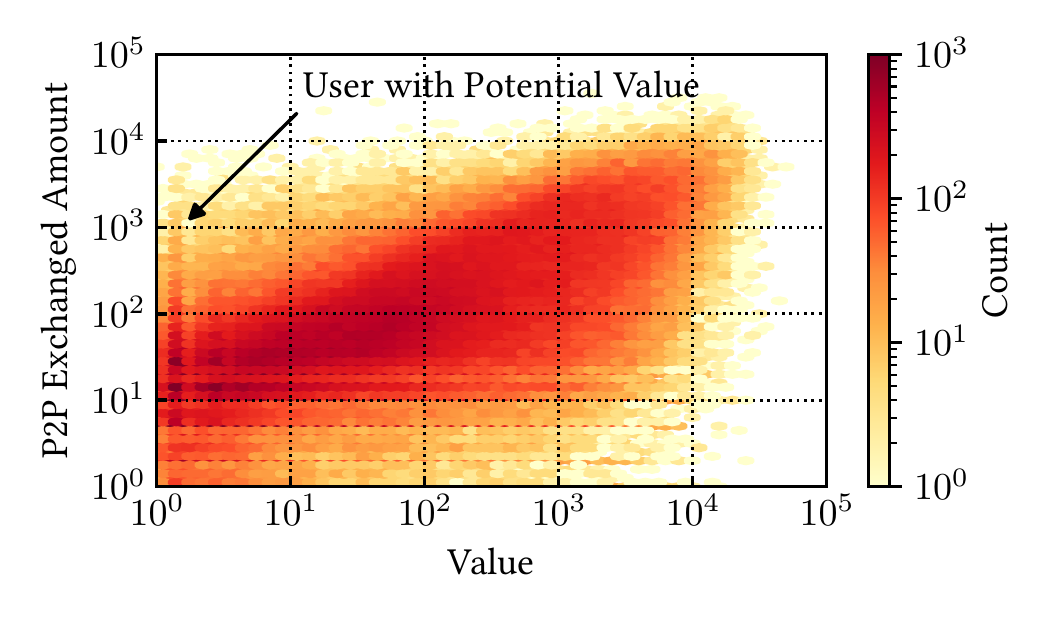}
    \centering
    \caption{Joint distribution of users' value and their total P2P transfer amount.}
    \label{fig:value-vs-p2p}
\end{figure}

On the prescriptive side, the novel understanding of the user base can be used to assist marketing decisions concerning incentive design and distribution. Concerning the distribution of incentives, by looking at the distribution of Value in Figure~\ref{fig:value-cdf}, we note that it is concentrated on a small portion of users. Consequently, flat investments mainly target users that produce small to null returns. This reveals the opportunity to design incentive distribution mechanisms based on targeting strategies. 
A simple and intuitive option is to target users with high value to incentivize them to spend or invite other valuable users. However, different strategies may be useful when the Value extracted from top users saturates. Indeed, users that already have good spending habits and a large network of peers may be unable to increase their performance even when given an incentive. Thus, an interesting strategy is to concentrate investments on users with high \emph{Potential Value}. Low Value users may be targeted based on their peer-to-peer activity (described by ``P2P'' relationships in our dataset). Despite being inactive on the P2B side, they still have high spending capabilities as they transfer a large amount of money. This type of marketing strategy is suggested and supported by our data. In Figure~\ref{fig:value-vs-p2p} we compare user Value with their P2P activity, measured by the amount of money they exchanged with other users. We observe a large portion of low Value users that are very active on the P2P side. They represent potentially valuable users that may bring great Value to the network if appropriately incentivized.
When it comes to implement targeted interventions, a central principle is that different users show different receptivity to incentives: the ability to target users by Value offers the possibility to tailor incentives to their characteristics.

%In summary, our methodology to measure user Value offers different possibilities, and the results and applications presented in this section represent only a limited selection of use cases that we have explored.

% \begin{table}[t]
% \centering
% \begin{tabular}{ |c|c|c|c| } 
%  \hline
%  & Intrinsic value & Network value & Value \\ 
%  \hline
%  Intrinsic value & 1.000 & 0.438 & 0.966 \\ 
%  \hline 
%  Network value & 0.438 & 1.000 & 0.558 \\ 
%  \hline
%  Value & 0.966 & 0.558 & 1.000 \\ 
%  \hline
% \end{tabular}
% \caption{Spearman correlation coefficients among value components.}
% \label{tab:spearman-corr}
% \end{table}

\subsection{Related Works}
\label{sec:related}

A number of previous works already proposed modeling payment systems in the form of a graph to achieve various goals. Already in 2008, Becher, Millard and Soramaki \cite{becher2008network} model the payment system CHAPS (a UK real-time gross settlement payment system) as a graph and investigated its properties. They found that it is a well-connected network whose properties hardly change from day to day. Soramaki, Bech, Arnold, Glass and Beyeler \cite{soramaki2007topology} perform a similar analysis for the interbank payments network. In the specific setting of mobile payment platforms \cite{aas2019evolution} models the evolution of Vipps, a peer-to-peer payment solution by Norway’s financial services group DNB, by means of a network formation process that combines preferential attachment with node fitness. Interestingly, both their setting and results are in the same spirit as ours, as they describe the evolution of the network as a process that is driven by a combination of network features and the intrinsic quality of users.

Recent works use graph mining approaches to detect fraud and prevent financial crime. Tam~\emph{et al.}~\cite{tam2019identifying} use Graph Convolution Network (GCN) to identify illicit accounts. They use GNN to learn embeddings of nodes and edges, which they exploit to characterize users and ultimately identify abnormal/suspicious financial transactions. Similarly, Li~\emph{et al.}~\cite{li2021temporal} model the payment system as a temporal interaction graph and propose a representation learning based framework to detect anomalies and illicit users. The applicability of these approaches in real-world systems is discussed by Kurshan and Shen \cite{kurshan2020graph}, who highlight the difficulties in implementing graph solutions in real-time financial transaction processing systems at an industrial scale.

Other problems have been addressed using graph approaches, including: Preventing deadlocks in payments~\cite{werman2018avoiding}; Resilience to disasters~\cite{pyrkina2019application}; Financial risk~\cite{martinez2014empirical, eboli2004systemic}; Privacy issues in ``I owe you (IOU)'' transaction networks~\cite{moreno2016listening}. A similar work to ours was carried out by Liu~\emph{et al.}~\cite{liu2019graph} from the Alipay payment platform. They propose a graph representation learning method for transaction networks that aims to optimize the allocation of incentives to run efficient marketing campaigns with a limited budget. In contrast to our approach, they aim to provide incentives to merchants rather than end users, and thus move towards modeling the sensitivity to incentives for each merchant.

\section{Conclusions}
\label{sec:conclu}

In this paper, we proposed an effective way to measure the value of users in modern payment platforms characterized by a network structure, and we demonstrated our approach on the YAP platform, an operational payment system with millions of users. Our methodology extends the classical RFM model used in marketing. It fits into the field of graph data science by leveraging the connections between entities to enrich the quantitative metrics of users with new knowledge derived from the interactions between users. Specifically, we defined a graph-based measure of user value that combines individual user information with the structure of the invitation network that describes the expansion of the platform. We provided a practical way to compute user value using an iterative graph algorithm based on the literature on opinion formation in social networks, with provable convergence on general topologies and finite time convergence in the case under study.

Our results confirmed the usefulness of our approach in analyzing and supporting the administration of payment platforms. We discussed relevant applications of our measure to inform business strategies. We first showed how our metric allows evaluating the performance of marketing campaigns, not only in terms of the number of newly acquired users, but also in terms of their value. We also illustrated how the proposed metric could suggest new targeting strategies that focus more on high value users or on low value users with high potential, allowing to adjust the trade-off between valorization of top users and exploration of the user base.

% Martino: sono indeciso se e cosa lasciare di questo paragrafo
%What we developed is a framework that can be shaped by business insights. Indeed, the model is highly flexible and there is control on a lot of aspects, such as what factors are more relevant in defining intrinsic value and what’s the relative importance of intrinsic and network value. Such flexibility requires close collaboration and a joint work with business experts, to both tune model parameter based on business knowledge and to identify most valuable applications, but at the same time this is what makes the model widely applicable.

%\section*{Acknowledgment}
%The research leading to these results has been funded by the SmartData@PoliTO center for Data Science technologies.

\bibliographystyle{IEEEtran}
\bibliography{references}

\appendices
\section{}
\label{sec:app}
%We provide in Appendix \ref{Appendix:a} details on the correspondance.
%We evolution of opinions and conventions in social networks
In the context of opinion formation and dynamic \cite{friedkin1990social}, it is assumed that the opinion of each individual in a networked society is locally influenced by that of their peers as each individual forms their own opinion by combining their personal belief with that of people in their neighborhood. 
We adapted these ideas to the novel setting of a payment platform so that the value of each user is obtained as a combination of their intrinsic value (referred as exogenous conditions) and the value of other users they are connected to by invitations (referred to as endogenous conditions). In particular, our model is a special case of the ``Group consensus models'' presented in~\cite[Section ``Model for static conditions'']{friedkin1990social} obtained by assuming that all parameters (including the graph structure) are immutable over time while the agents' opinion is the only factor to evolve and by making the following identifications:
\begin{itemize}
    \item vector $Y$, which represents the opinion of agents in \cite{friedkin1990social}, corresponds to vector $V$ measuring the value of users,
    \item the contribution of exogenous variables, denoted by $XB$ in \cite{friedkin1990social}, corresponds to the vector of intrinsic user values~$I$,
    \item the scalar value $\alpha$ weights the relevance of connections in both approaches,
    \item the scalar value $\beta$, which in \cite{friedkin1990social} weights the effect of exogenous conditions, is set to $1$ in our setting.
\end{itemize}
This correspondence also guarantees convergence of our algorithm \eqref{eq:algo-iteration} to the solution of \eqref{eq:value-measure} for generic graph topologies, as this is equivalent to convergence of the opinion dynamics process proved in~\cite{friedkin1990social}.

\end{document}